\documentclass[twocolumn,letterpaper,aps,prd,longbibliography,superscriptaddress,nofootinbib,floatfix]{revtex4-1}

\usepackage{graphicx}	
\usepackage{xspace}	
\usepackage{amsmath}

\newcommand{\pt}{\mbox{$p_T$}\xspace}

\newcommand{\sqs}{\mbox{$\sqrt{s}$}\xspace}

\newcommand{\pp}{\mbox{$p$$+$$p$}\xspace}

\newcommand{\pout}{\mbox{$p_{\rm out}$}\xspace}
\newcommand{\pttrig}{\mbox{$p_T^{\rm trig}$}\xspace}
\newcommand{\rgammaiso}{\mbox{$R_\gamma^{\rm iso}$}\xspace}
\newcommand{\dphi}{\mbox{$\Delta\phi$}\xspace}

\newcommand{\pion}{\mbox{$\pi^0$}\xspace}

\newcommand{\mapfxn}{\mbox{$P(p_T^{\pi^0},p_T^\gamma)$}\xspace}
\newcommand{\ptassoc}{\mbox{$p_{T}^{\rm assoc}$}\xspace}
\newcommand{\rgamma}{\mbox{$R_\gamma$}\xspace}

\newcommand{\jt}{\mbox{$\sqrt{\langle j_T^2\rangle}$}\xspace}
\newcommand{\xe}{\mbox{$x_E$}\xspace}
\newcommand{\gevc}{\mbox{GeV/$c$}\xspace}

\begin{document}

\title{Nonperturbative transverse-momentum-dependent effects in dihadron 
and direct photon-hadron angular correlations in $p$$+$$p$ 
collisions at $\sqrt{s}=200$ GeV}

\newcommand{\abilene}{Abilene Christian University, Abilene, Texas 79699, USA}
\newcommand{\augie}{Department of Physics, Augustana University, Sioux Falls, South Dakota 57197, USA}
\newcommand{\banaras}{Department of Physics, Banaras Hindu University, Varanasi 221005, India}
\newcommand{\barc}{Bhabha Atomic Research Centre, Bombay 400 085, India}
\newcommand{\baruch}{Baruch College, City University of New York, New York, New York, 10010 USA}
\newcommand{\bnlcoll}{Collider-Accelerator Department, Brookhaven National Laboratory, Upton, New York 11973-5000, USA}
\newcommand{\bnlphys}{Physics Department, Brookhaven National Laboratory, Upton, New York 11973-5000, USA}
\newcommand{\caucr}{University of California-Riverside, Riverside, California 92521, USA}
\newcommand{\charlesczech}{Charles University, Ovocn\'{y} trh 5, Praha 1, 116 36, Prague, Czech Republic}
\newcommand{\chonbuk}{Chonbuk National University, Jeonju, 561-756, Korea}
\newcommand{\cns}{Center for Nuclear Study, Graduate School of Science, University of Tokyo, 7-3-1 Hongo, Bunkyo, Tokyo 113-0033, Japan}
\newcommand{\colorado}{University of Colorado, Boulder, Colorado 80309, USA}
\newcommand{\columbia}{Columbia University, New York, New York 10027 and Nevis Laboratories, Irvington, New York 10533, USA}
\newcommand{\czechtech}{Czech Technical University, Zikova 4, 166 36 Prague 6, Czech Republic}
\newcommand{\debrecen}{Debrecen University, H-4010 Debrecen, Egyetem t{\'e}r 1, Hungary}
\newcommand{\elte}{ELTE, E{\"o}tv{\"o}s Lor{\'a}nd University, H-1117 Budapest, P{\'a}zm{\'a}ny P.~s.~1/A, Hungary}
\newcommand{\eszterhazy}{Eszterh\'azy K\'aroly University, K\'aroly R\'obert Campus, H-3200 Gy\"ongy\"os, M\'atrai \'ut 36, Hungary}
\newcommand{\ewha}{Ewha Womans University, Seoul 120-750, Korea}
\newcommand{\fsu}{Florida State University, Tallahassee, Florida 32306, USA}
\newcommand{\gsu}{Georgia State University, Atlanta, Georgia 30303, USA}
\newcommand{\hiroshima}{Hiroshima University, Kagamiyama, Higashi-Hiroshima 739-8526, Japan}
\newcommand{\howard}{Department of Physics and Astronomy, Howard University, Washington, DC 20059, USA}
\newcommand{\ihepprot}{IHEP Protvino, State Research Center of Russian Federation, Institute for High Energy Physics, Protvino, 142281, Russia}
\newcommand{\illuiuc}{University of Illinois at Urbana-Champaign, Urbana, Illinois 61801, USA}
\newcommand{\inrras}{Institute for Nuclear Research of the Russian Academy of Sciences, prospekt 60-letiya Oktyabrya 7a, Moscow 117312, Russia}
\newcommand{\instpasczech}{Institute of Physics, Academy of Sciences of the Czech Republic, Na Slovance 2, 182 21 Prague 8, Czech Republic}
\newcommand{\isu}{Iowa State University, Ames, Iowa 50011, USA}
\newcommand{\jaea}{Advanced Science Research Center, Japan Atomic Energy Agency, 2-4 Shirakata Shirane, Tokai-mura, Naka-gun, Ibaraki-ken 319-1195, Japan}
\newcommand{\kek}{KEK, High Energy Accelerator Research Organization, Tsukuba, Ibaraki 305-0801, Japan}
\newcommand{\korea}{Korea University, Seoul, 136-701, Korea}
\newcommand{\kurchatov}{National Research Center ``Kurchatov Institute", Moscow, 123098 Russia}
\newcommand{\kyoto}{Kyoto University, Kyoto 606-8502, Japan}
\newcommand{\lawllnl}{Lawrence Livermore National Laboratory, Livermore, California 94550, USA}
\newcommand{\losalamos}{Los Alamos National Laboratory, Los Alamos, New Mexico 87545, USA}
\newcommand{\lund}{Department of Physics, Lund University, Box 118, SE-221 00 Lund, Sweden}
\newcommand{\lyon}{IPNL, CNRS/IN2P3, Univ Lyon, Université Lyon 1, F-69622, Villeurbanne, France}
\newcommand{\maryland}{University of Maryland, College Park, Maryland 20742, USA}
\newcommand{\mass}{Department of Physics, University of Massachusetts, Amherst, Massachusetts 01003-9337, USA}
\newcommand{\michigan}{Department of Physics, University of Michigan, Ann Arbor, Michigan 48109-1040, USA}
\newcommand{\muhlenberg}{Muhlenberg College, Allentown, Pennsylvania 18104-5586, USA}
\newcommand{\nara}{Nara Women's University, Kita-uoya Nishi-machi Nara 630-8506, Japan}
\newcommand{\natmephi}{National Research Nuclear University, MEPhI, Moscow Engineering Physics Institute, Moscow, 115409, Russia}
\newcommand{\newmex}{University of New Mexico, Albuquerque, New Mexico 87131, USA}
\newcommand{\nmsu}{New Mexico State University, Las Cruces, New Mexico 88003, USA}
\newcommand{\ohio}{Department of Physics and Astronomy, Ohio University, Athens, Ohio 45701, USA}
\newcommand{\ornl}{Oak Ridge National Laboratory, Oak Ridge, Tennessee 37831, USA}
\newcommand{\orsay}{IPN-Orsay, Univ.~Paris-Sud, CNRS/IN2P3, Universit\'e Paris-Saclay, BP1, F-91406, Orsay, France}
\newcommand{\peking}{Peking University, Beijing 100871, People's Republic of China}
\newcommand{\pnpi}{PNPI, Petersburg Nuclear Physics Institute, Gatchina, Leningrad region, 188300, Russia}
\newcommand{\riken}{RIKEN Nishina Center for Accelerator-Based Science, Wako, Saitama 351-0198, Japan}
\newcommand{\rikjrbrc}{RIKEN BNL Research Center, Brookhaven National Laboratory, Upton, New York 11973-5000, USA}
\newcommand{\rikkyo}{Physics Department, Rikkyo University, 3-34-1 Nishi-Ikebukuro, Toshima, Tokyo 171-8501, Japan}
\newcommand{\saispbstu}{Saint Petersburg State Polytechnic University, St.~Petersburg, 195251 Russia}
\newcommand{\seoulnat}{Department of Physics and Astronomy, Seoul National University, Seoul 151-742, Korea}
\newcommand{\stonybrkc}{Chemistry Department, Stony Brook University, SUNY, Stony Brook, New York 11794-3400, USA}
\newcommand{\stonycrkp}{Department of Physics and Astronomy, Stony Brook University, SUNY, Stony Brook, New York 11794-3800, USA}
\newcommand{\tenn}{University of Tennessee, Knoxville, Tennessee 37996, USA}
\newcommand{\titech}{Department of Physics, Tokyo Institute of Technology, Oh-okayama, Meguro, Tokyo 152-8551, Japan}
\newcommand{\tsukuba}{Tomonaga Center for the History of the Universe, University of Tsukuba, Tsukuba, Ibaraki 305, Japan}
\newcommand{\vandy}{Vanderbilt University, Nashville, Tennessee 37235, USA}
\newcommand{\weizmann}{Weizmann Institute, Rehovot 76100, Israel}
\newcommand{\wigner}{Institute for Particle and Nuclear Physics, Wigner Research Centre for Physics, Hungarian Academy of Sciences (Wigner RCP, RMKI) H-1525 Budapest 114, POBox 49, Budapest, Hungary}
\newcommand{\yonsei}{Yonsei University, IPAP, Seoul 120-749, Korea}
\newcommand{\zagreb}{Department of Physics, Faculty of Science, University of Zagreb, Bijeni\v{c}ka c.~32 HR-10002 Zagreb, Croatia}
\affiliation{\abilene}
\affiliation{\augie}
\affiliation{\banaras}
\affiliation{\barc}
\affiliation{\baruch}
\affiliation{\bnlcoll}
\affiliation{\bnlphys}
\affiliation{\caucr}
\affiliation{\charlesczech}
\affiliation{\chonbuk}
\affiliation{\cns}
\affiliation{\colorado}
\affiliation{\columbia}
\affiliation{\czechtech}
\affiliation{\debrecen}
\affiliation{\elte}
\affiliation{\eszterhazy}
\affiliation{\ewha}
\affiliation{\fsu}
\affiliation{\gsu}
\affiliation{\hiroshima}
\affiliation{\howard}
\affiliation{\ihepprot}
\affiliation{\illuiuc}
\affiliation{\inrras}
\affiliation{\instpasczech}
\affiliation{\isu}
\affiliation{\jaea}
\affiliation{\kek}
\affiliation{\korea}
\affiliation{\kurchatov}
\affiliation{\kyoto}
\affiliation{\lawllnl}
\affiliation{\losalamos}
\affiliation{\lund}
\affiliation{\lyon}
\affiliation{\maryland}
\affiliation{\mass}
\affiliation{\michigan}
\affiliation{\muhlenberg}
\affiliation{\nara}
\affiliation{\natmephi}
\affiliation{\newmex}
\affiliation{\nmsu}
\affiliation{\ohio}
\affiliation{\ornl}
\affiliation{\orsay}
\affiliation{\peking}
\affiliation{\pnpi}
\affiliation{\riken}
\affiliation{\rikjrbrc}
\affiliation{\rikkyo}
\affiliation{\saispbstu}
\affiliation{\seoulnat}
\affiliation{\stonybrkc}
\affiliation{\stonycrkp}
\affiliation{\tenn}
\affiliation{\titech}
\affiliation{\tsukuba}
\affiliation{\vandy}
\affiliation{\weizmann}
\affiliation{\wigner}
\affiliation{\yonsei}
\affiliation{\zagreb}
\author{C.~Aidala} \affiliation{\michigan} 
\author{Y.~Akiba} \email[PHENIX Spokesperson: ]{akiba@rcf.rhic.bnl.gov} \affiliation{\riken} \affiliation{\rikjrbrc} 
\author{M.~Alfred} \affiliation{\howard} 
\author{V.~Andrieux} \affiliation{\michigan} 
\author{N.~Apadula} \affiliation{\isu} 
\author{H.~Asano} \affiliation{\kyoto} \affiliation{\riken} 
\author{B.~Azmoun} \affiliation{\bnlphys} 
\author{V.~Babintsev} \affiliation{\ihepprot} 
\author{A.~Bagoly} \affiliation{\elte} 
\author{N.S.~Bandara} \affiliation{\mass} 
% ...
\collaboration{PHENIX Collaboration}  \noaffiliation

\date{\today}


\begin{abstract}


Dihadron and isolated direct photon-hadron angular correlations 
are measured in $p$$+$$p$ collisions at $\sqrt{s}=200$ GeV. The 
correlations are sensitive to nonperturbative initial-state and 
final-state transverse momentum $k_T$ and $j_T$ in the azimuthal 
nearly back-to-back region $\Delta\phi\sim\pi$. In this region, 
transverse-momentum-dependent evolution can be studied when 
several different hard scales are measured. To have sensitivity to 
small transverse momentum scales, nonperturbative momentum widths 
of $p_{\rm out}$, the out-of-plane transverse momentum component 
perpendicular to the trigger particle, are measured. These widths 
are used to investigate possible effects from 
transverse-momentum-dependent factorization breaking. When 
accounting for the longitudinal momentum fraction of the away-side 
hadron with respect to the near-side trigger particle, the widths 
are found to increase with the hard scale; this is qualitatively 
similar to the observed behavior in Drell-Yan and semi-inclusive 
deep-inelastic scattering interactions. The momentum widths are 
also studied as a function of center-of-mass energy by comparing 
to previous measurements at $\sqrt{s}=510$ GeV. The 
nonperturbative jet widths also appear to increase with $\sqrt{s}$ 
at a similar $x_T$, which is qualitatively consistent to similar 
measurements in Drell-Yan interactions. To quantify the magnitude 
of any transverse-momentum-dependent factorization breaking 
effects, calculations will need to be performed to compare to 
these measurements.
 
\end{abstract}

	



\maketitle

\section{Introduction}

Quantum chromodynamics (QCD) research has entered a period in which the 
focus of nucleon structure has shifted from a one-dimensional to a 
multidimensional picture. To take into account additional degrees of 
freedom besides the longitudinal momentum of partons within hadrons, the 
transverse-momentum-dependent (TMD) framework has been developed which 
also accounts for partons' transverse momentum. Rather than parton 
distribution functions (PDFs) and fragmentation functions (FFs) being 
integrated over transverse momentum degrees of freedom, TMD PDFs and TMD 
FFs contain explicit dependence on the nonperturbative transverse 
momentum of the parton within the nucleon. This dependence offers a way 
to describe the three dimensional momentum distribution of unpolarized 
partons within unpolarized hadrons, in addition to a variety of 
spin-spin and spin-momentum correlations when the parton and nucleon 
spin states are considered. 

In the past decade, significant effort has been placed on measuring 
asymmetries that are sensitive to PDFs and FFs within the TMD framework. 
Semi-inclusive deep-inelastic scattering (SIDIS) and Drell-Yan (DY) 
measurements have shown empirical evidence for nonzero spin momentum 
correlations in the initial 
state~\cite{Airapetian:2004tw,Adolph:2016dvl,Huang:2011bc,Zhu:2008sj,Guanziroli:1987rp}. 
Additionally, $e^+e^-$ annihilation and SIDIS measurements have shown 
nonzero spin-momentum correlations in the final-state hadronization 
process~\cite{Abe:2005zx,TheBABAR:2013yha,Adolph:2014zba}. With the 
development of robust and theoretically interpretable jet finding 
algorithms, there have also been several measurements studying 
nonperturbative hadronization in unpolarized \pp 
collisions~\cite{Aad:2011sc,ALICE:2014dla} as well as spin-momentum 
correlations in polarized \pp 
collisions~\cite{Adamczyk:2017wld,Adamczyk:2017ynk}. Transverse single 
spin asymmetries of inclusive hadron production in \pp collisions of up 
to $\sim$40\% also indicate large spin-momentum correlations within the 
nucleon and/or the process of hadronization~\cite{Aidala:2012mv}; 
however, these measurements cannot probe functions within the TMD 
framework directly because there is not a simultaneous measurement of a 
small and hard transverse momentum scale. \par

The focus on multidimensional parton structure has brought the 
importance of soft gluon exchanges in hard interactions to the forefront 
of QCD research. In particular, the role of color exchanges due to soft 
gluon interactions with remnants of the hard scattering has brought to 
light fundamental predictions about QCD as a gauge-invariant quantum 
field theory. For example, the Sivers TMD 
PDF~\cite{Sivers:1989cc,Sivers:1990fh}, which correlates the partonic 
transverse momentum with the nucleon spin, is predicted to exhibit 
modified universality when measured in the SIDIS and DY 
processes~\cite{Collins:2002kn}. The underlying physical cause of this 
prediction is the interference between color fields in the two processes 
when an initial-state gluon is exchanged in DY vs.~a final-state gluon 
in SIDIS. The Sivers function has already been measured to be nonzero in 
SIDIS~\cite{Airapetian:2004tw}. The first measurements of the DY-like W 
boson and DY transverse single spin asymmetry were recently 
reported~\cite{Adamczyk:2015gyk,Aghasyan:2017jop}, and are consistent 
with the predicted sign change of the Sivers function. However, the 
uncertainties on the measurements are still large enough that a 
definitive conclusion cannot be drawn; more data will ultimately be 
necessary as the prediction is for the entire Sivers distribution as a 
function of the partonic longitudinal and transverse momentum $x$ and 
$k_T$.

In leading order perturbative QCD processes where a colored parton is 
exchanged in the hard interaction, and thus color is necessarily present 
in both the initial and final states, soft gluon exchanges can lead to 
new effects in a TMD framework similarly to the predicted modified 
universality of certain TMD PDFs. In hadronic collisions where a 
final-state hadron is measured and the process is sensitive to a small 
transverse momentum scale, factorization breaking has been 
predicted~\cite{Bomhof:2006dp,Collins:2007jp,Collins:2007nk,Rogers:2010dm}. 
These processes can offer sensitivity to the non-Abelian nature of QCD. 
In processes where factorization is broken, the nonperturbative objects 
can no longer be factorized into a convolution of TMD PDFs and TMD FFs 
due to the complex color flows that are possible throughout the hard 
scattering and remnants of the collision and thus connect the initial 
and final state hadrons. In both cases of factorization breaking and 
modified universality of certain TMD PDFs, gluon exchanges with the 
remnants cannot be eliminated via gauge transformations. There have also 
been recent studies showing that factorization is broken for certain 
processes even at the collinear multi-loop 
level~\cite{Catani:2011st,Forshaw:2012bi}; however, the focus of this 
work is to probe TMD factorization breaking effects.

In the past decade, the role of color in hadronic interactions has also 
been explored in several different types of observables in \pp 
collisions. The measurement of collective behavior in high multiplicity 
\pp collisions has prompted new studies of global observables in 
addition to interference effects between necessarily color connected 
multiple partonic 
interactions~\cite{Abelev:2013bla,Ortiz:2013yxa,Blok:2017pui}. 
Measurements of dijet+jet and direct photon-jet+jet correlations at the 
Large Hadron Collider studying color-coherence effects have found 
effects from color-radiation patterns specific to hadronic 
collisions~\cite{Chatrchyan:2013fha,Aaboud:2016sdm}. New dijet 
observables which rely on jet substructure have been shown to be 
sensitive to color flow in $t\bar{t}$ events~\cite{Aad:2015lxa}. With 
theoretical advances in jet substructure techniques that may be more 
sensitive to factorization breaking effects, such as within 
soft-collinear effective theory~\cite{Schwartz:2018obd}, future 
measurements involving jet substructure may be an effective way to probe 
soft radiation patterns that allow quantification of color flow effects 
similarly to Ref.~\cite{Aad:2015lxa}. The advancement of both 
experimental and theoretical techniques has allowed the magnitude of 
effects from color flow and color connections to be probed in a variety 
of ways across various subfields of QCD.

A previous measurement~\cite{Adare:2016bug} in \pp collisions at 
\sqs=~510 GeV found that nonperturbative momentum widths in dihadron 
and direct photon-hadron angular correlations binned in a fixed range 
of the away-side associated hadron \pt decreased with the hard scale 
of the interaction, which is qualitatively opposite to what is 
expected from Collins-Soper-Sterman (CSS) 
evolution~\cite{Adare:2016bug} which comes directly out of the 
derivation of TMD factorization~\cite{Collins:2012ss}. This study is 
intended to extend the analysis of Ref.~\cite{Adare:2016bug} for 
processes which are predicted to violate factorization; however, this 
measurement is made at \sqs=~200 GeV.  It supplants the previous 
PHENIX analysis at \sqs=~200 GeV~\cite{Adare:2010yw} with an 
approximately four times increase in integrated luminosity. 
The two different center-of-mass energies of Ref.~\cite{Adare:2016bug} 
and this analysis allow TMD effects to be studied as a function of 
\sqs in processes predicted to violate factorization.

\section{Experiment Details}

\begin{figure}[tbh]
	\includegraphics[width=1\linewidth]{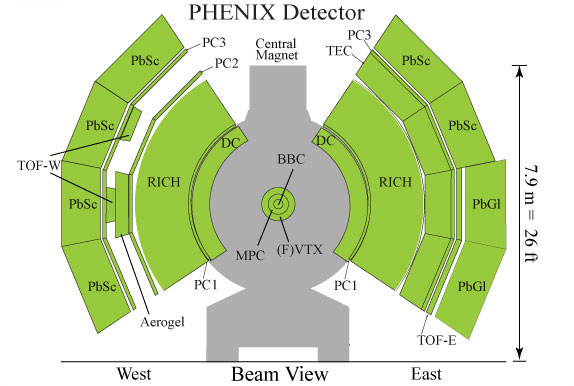}
    \caption{A cross sectional view along the beam line shows the two 
central arms of the PHENIX detector.}
   	\label{fig:phenix}
\end{figure}

In 2015 the PHENIX experiment collected data from \pp collisions at 
\sqs=~200 GeV. A total minimum bias integrated luminosity of 60 
pb$^{-1}$ was used for the analysis of dihadron and direct photon-hadron 
angular correlations, from which data quality assurance and $z$ vertex 
cuts of $|z_{\rm vtx}|<$~30 cm were applied. The PHENIX detector measures 
two particle angular correlations via its electromagnetic calorimeter 
(EMCal) and drift chamber (DC) plus pad chamber (PC) tracking system 
located in two central-rapidity arms. The central arms are nearly 
back-to-back in azimuth and each cover approximately 
$\Delta\phi\sim\pi/2$ radians in azimuthal angle and 
$\Delta\eta\sim$~0.7 units in pseudorapidity centered about midrapidity. 
A schematic diagram of the two central arms is shown in 
Fig.~\ref{fig:phenix}. Detailed descriptions of the PHENIX central arm 
spectrometers can be found in 
Refs.~\cite{Adcox:2003zp,Aphecetche:2003zr}

The PHENIX EMCal~\cite{Aphecetche:2003zr} comprises eight sectors, 
four in 
each arm.  Six of the sectors are lead-scintillator sampling 
calorimeters, and the other two are lead-glass \v{C}erenkov calorimeters. A 
high-energy-photon trigger in the EMCal is used to identify events with 
a high-\pt photon. Photons are identified with a shower shape cut that 
removes charged hadrons as well as most high energy clusters that 
overlap closely with another photon, which helps eliminate $\pion$ 
merging effects at high \pt. The granularity of the EMCal as well as the 
shower shape cuts allow for \pion and $\eta$ reconstruction up to 
$\sim$20 GeV in the diphoton channel, which allows for direct photon 
measurements as well. In this analysis isolated photons are measured 
between $5<\pt<15$ \gevc, and neutral pions are measured between 
$4<p_T<15$ \gevc. Previous \pion, $\eta$, and direct photon cross 
sections as well as \pion-hadron or direct photon-hadron correlation can 
be found in 
Refs.~\cite{Adare:2012yt,Adare:2008qb,Adare:2010cy,Adare:2010yw,Adare:2016bug}.

The PHENIX tracking system~\cite{Adcox:2003zp} measures nonidentified 
charged hadrons with a DC and PC tracking system. The PC1, located 
radially behind the drift chamber, and PC3, located radially in front of 
the EMCal, allow for tracks identified in the DC to be matched with PC 
hits. The track matching condition in the PC1 and especially the PC3 
reduces secondary tracks from conversions or decays. The ring-imaging 
\v{C}erenkov system, located radially behind the DC in 
Fig.~\ref{fig:phenix}, is also used to remove electrons from the charged 
hadron sample. The tracking system is also used to suppress hadronic 
shower contamination when identifying photons by matching tracks in the 
DC and PC to showers in the EMCal and subsequently removing them. In 
this analysis charged hadrons are collected from $0.5<p_T<10$ \gevc. A 
previous nonidentified charged hadron cross section from PHENIX can be 
found in e.g. Ref.~\cite{Adler:2005in}.

\section{Analysis}

Dihadron and direct photon-hadron correlation functions are constructed 
following the methods of Refs.~\cite{Adare:2010yw,Adare:2016bug}. The 
correlation functions are constructed by measuring the raw number of 
correlated trigger particles and associated hadrons, where the trigger 
particle refers to either a leading \pion or isolated photon. To account 
for the PHENIX acceptance, the yields are corrected by a mixed-event 
distribution by constructing correlations between trigger particles from 
one event and associated hadrons from a different event. The mixed event 
distribution is constructed on a run-by-run basis to quantify any 
changing efficiencies of the detector with time. The correlation 
functions are additionally corrected by a charged hadron efficiency to 
quantify the inefficiency of the DC and PC tracking system. The charged 
hadron efficiency is determined by simulating single particle hadrons in 
a {\sc{GEANT3}}-based description of the PHENIX 
detector~\cite{Adler:2003zv}. After corrections the correlation 
functions are divided by the total number of trigger particles to 
construct a per-trigger yield. In this analysis a 9\% uncertainty is 
assigned to the charged hadron efficiency, which amounts to an overall 
normalization uncertainty on the per-trigger yields. In total the 
correlation functions correspond to yields within full azimuthal 
coverage and $|\eta|<0.35$. The correlations are constructed similarly 
to previous PHENIX analyses in 
Refs.~\cite{Adare:2009aa,Adler:2006sc,Adare:2009vd,Adare:2010yw,Adare:2016bug}.

\subsection{Statistical subtraction of decay photons}

The correlation functions can be constructed for any generic 
trigger-associated hadron combination. To identify direct photon-hadron 
correlations, an additional statistical subtraction must be applied to 
account for the background due to decay photon-hadron correlations. 
Reference~\cite{Adare:2009vd} used a method which statistically 
subtracts the decay photon-hadron per-trigger yield component from a 
total inclusive photon-hadron per-trigger yield component with the 
following equation

\begin{equation}\label{eq:rgamma_eq}
	Y_{\rm direct} = \frac{1}{\rgamma-1}(\rgamma Y_{\rm inclusive}-Y_{\rm decay})
\end{equation}

\noindent In this equation Y is the per-trigger yield of a particular 
component denoted by the subscript, and \rgamma is the relative 
contribution of direct photons to decay photons defined as $\rgamma 
\equiv N_{\rm inclusive}/N_{\rm decay}$. In Ref.~\cite{Adare:2009vd}, 
direct photons are defined as any photon not from a decay process, and 
includes next-to-leading order fragmentation photons.

To reduce the presence of fragmentation photons, 
Refs.~\cite{Adare:2010yw,Adare:2016bug} implemented an isolation cone 
criterion in the method, which we also use in this analysis. This has 
the added benefit of also reducing the decay photon background and thus 
providing a larger signal to background ratio for the direct photons. 
Additionally, photons that could be tagged as coming from \pion or 
$\eta$ decays were removed. Because the tagging procedure does not remove 
all decay photons, Eq.~\ref{eq:rgamma_eq} was modified to include these 
cuts and introduce isolated photon quantities

\begin{equation}
	Y_{\rm direct}^{\rm iso} = \frac{1}{\rgammaiso-1}(\rgammaiso Y_{\rm inclusive}^{\rm iso}-Y_{\rm decay}^{\rm iso})
\end{equation}

\noindent where again the trigger particles are noted in the subscripts for the 
various per-trigger yields and ``iso'' refers to both ``isolated'' and 
not tagged as coming from a \pion decay. Here \rgammaiso is the relative 
contribution of isolated direct photons to isolated decay photons, 
defined in a similar way to \rgamma except with isolated quantities. 
While the subtraction procedure of Ref.~\cite{Adare:2009vd} removes all 
decay photons, the modified isolated subtraction procedure removes 
background due to isolated decay photon-hadron correlations, which, in 
the PHENIX acceptance, are due most often to isolated neutral pion 
decays which decay asymmetrically such that the low-\pt photon is not 
detected.

To implement the isolated statistical subtraction procedure, isolation 
and tagging cuts are applied at the event-by-event level in the data 
analysis. Inclusive photon candidates are removed from the analysis if a 
partner photon with \pt$>$~500 MeV/$c$ is found such that the invariant 
mass of the pair falls within the \pion or $\eta$ invariant mass 
regions, $0.118<m_{\gamma\gamma}<0.162$ and $0.5<m_{\gamma\gamma}<0.6$ 
GeV/$c^2$. In addition to the tagging cuts, an isolation cut is 
implemented. The isolation criterion is that the sum of EMCal energy and 
\pt of charged tracks within a cone radius of 0.3 radians must be less 
than 10\% of the candidate photons energy, similar to 
Ref.~\cite{Adare:2010yw}. To reduce the effects of the detector 
acceptance, candidate isolated photons are also required to be $\sim$0.1 
radians from the edge of the detector in both $\phi$ and $\eta$. This 
forces a large portion of the cone to fall within the PHENIX acceptance.

The number of isolated decay photon-hadron pairs are not known {\it{a 
priori}} because an unknown fraction of the isolated photons still come 
from decay processes. The isolated decay photon-hadron correlations are 
determined with a Monte Carlo generated probability density function 
that quantifies the probability of an isolated \pion to decay to an 
isolated photon within the PHENIX acceptance such that the photon was 
not able to be tagged as coming from a decay process. The functions are 
used to map the isolated \pion-hadron correlations to the corresponding 
daughter isolated photon-hadron correlations for use in the statistical 
subtraction procedure, where the isolation criterion for the \pion meson 
is the same as described above for single photons. In total a 4\% 
systematic uncertainty was assigned for the statistical subtraction 
process, which includes additional background coming from higher mass 
state decays. With the probability functions, the decay photon-hadron 
per-trigger yields can be determined by

\begin{equation}
Y_{\rm decay}^{\rm iso} = \frac{\sum_{N^{\rm iso}_{\pi-h}}\mapfxn N_{\pi^0-h}^{\rm iso}}{\sum_{N^{\rm iso}_\pi}\mapfxn N_{\pi^0}^{\rm iso}}
\end{equation}

\noindent where \mapfxn is the probability density function described 
above and $N$ are the number of isolated \pion-hadron pairs or isolated 
\pion triggers measured as indicated in the subscripts.

To determine \rgammaiso, the quantity \rgamma must be corrected for the 
isolation and tagging cuts. The \rgamma was previously measured in 
Ref.~\cite{Adare:2010yw}, so these values are used and corrected with 
the measured isolation and tagging efficiencies of this analysis. The 
quantity is the ratio of isolated inclusive photons to isolated decay 
photons, is dependent only on the photon $p_T$, and can be written as

\begin{equation}\label{eq:rgammaiso_eq}
\begin{split}
\rgammaiso(p_T^\gamma) & = \frac{N_{\rm inclusive}^{\rm iso}}{N_{\rm decay}^{\rm iso}} \\ 
& =  \frac{R_\gamma}{(1-\epsilon_{\rm decay}^{\rm tag})
(1-\epsilon_{\rm decay}^{\rm niso})}\times \\  & \frac{N_{\rm inclusive}
-N_{\rm decay}^{\rm tag}-N_{\rm inclusive}^{\rm niso}}{N_{\rm inclusive}}
\end{split}
\end{equation}

\noindent where ``niso'' refers to ``not isolated,'' $\epsilon_{\rm 
decay}^{\rm tag}$ is the tagging efficiency, $\epsilon_{\rm decay}^{\rm 
niso}$ is the isolation efficiency, and the various $N$ values are the 
number of photons that correspond to the subscript and superscript with 
which they are associated. The bottom-most fraction in 
Eq.~\ref{eq:rgammaiso_eq} is simply the number of isolated inclusive 
photons divided by the total number of inclusive photons and can be 
determined by just counting the photons that pass the necessary cuts. 
The decay photon tagging efficiency can be written as $\epsilon_{\rm 
decay}^{\rm tag} = \rgamma N_{\rm decay}^{\rm tag}/N_{\rm inclusive}$ 
and can also be determined by counting the number of photons that pass 
the cuts. To determine the isolation efficiency, the isolation cut is 
applied at the level of the parent \pion and the decay probability 
functions are used to map the effect to the daughter photons. The 
isolation efficiency can be written as

\begin{equation}\label{eq:isoeff_eq}
\epsilon^{\rm niso}_{\rm decay} = \left(1+\frac{\sum_\pi\mapfxn\cdot N_{\pi}^{\rm iso}}{\sum_\pi\mapfxn\cdot N_{\pi}^{\rm niso}}\right)^{-1}
\end{equation}

\begin{figure}[tbh]
    \includegraphics[width=1\linewidth]{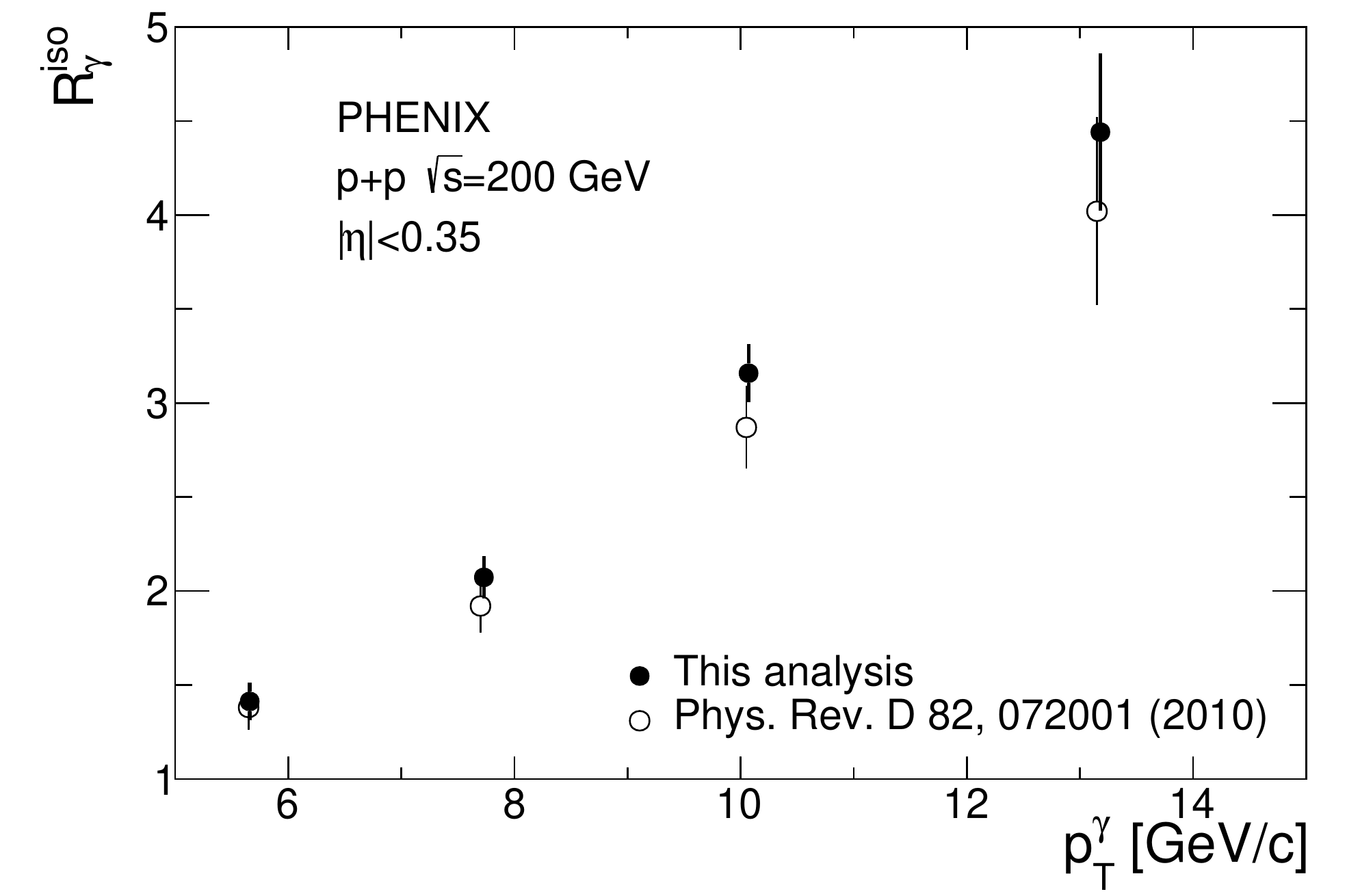}
   	\caption{Measured \rgammaiso in this analysis compared to 
Ref.~\cite{Adare:2010yw}. The statistic and systematic uncertainties are 
added in quadrature and are dominated by the systematic uncertainties.}
    \label{fig:rgammaiso}
\end{figure}

\noindent Therefore, all of the necessary components can be determined 
to calculate \rgammaiso. Figure~\ref{fig:rgammaiso} shows the measured 
values of \rgammaiso as a function of $p_T^\gamma$ in this analysis 
compared to the previous PHENIX publication~\cite{Adare:2010yw} at the 
same \sqs. The values show consistency with the previous analysis and 
are all larger than unity, indicating the signal-to-background of the 
isolated direct photons to decay photons. These can also be compared to 
the values of \rgamma in Refs.~\cite{Adare:2010yw,Adare:2009vd} showing 
that the tagging and isolation cuts increase the signal-to-background.

\section{Results}

\begin{figure*}[tbh]
\includegraphics[width=1\textwidth]{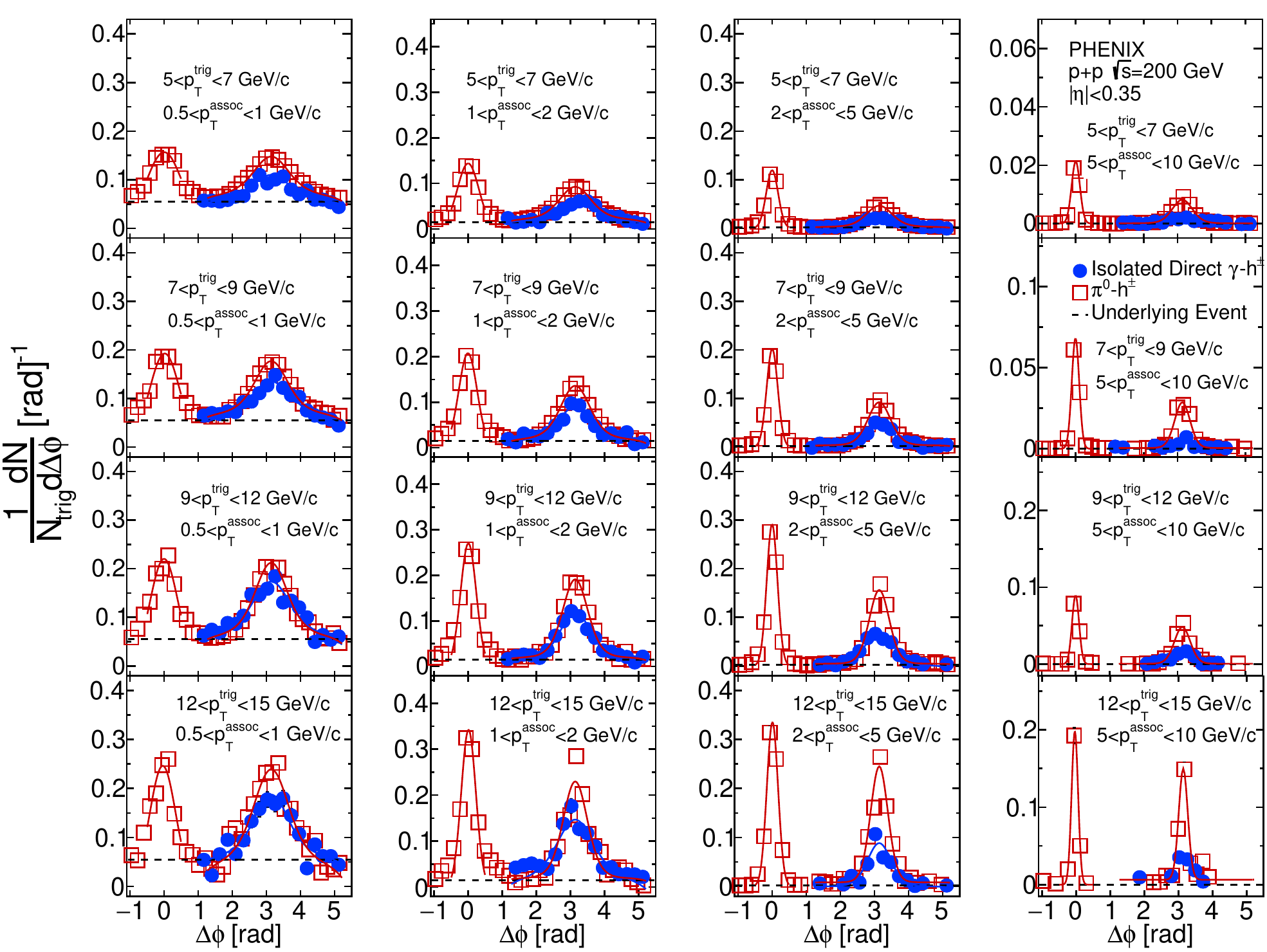}
\caption{The per-trigger yields are shown as a function of \dphi in 
several \pttrig$\otimes~$\ptassoc bins. The black dashed line shows an 
estimation of the underlying event yield to more clearly show the 
away-side jet yield. The 9\% charged hadron normalization uncertainty is 
not explicitly shown on the figure.}
\label{fig:deltaphi_correlations}
\end{figure*}

\subsection{Azimuthal correlations}

Figure~\ref{fig:deltaphi_correlations} shows the correlation functions 
measured in various \pttrig and \ptassoc bins. In the figure the red 
open points are the dihadron correlations, while the filled blue points 
are the isolated direct photon-hadron correlations. A black dashed line 
estimates the underlying event pedestal to emphasize the jet yields. The 
per-trigger yields as a function of \dphi show the expected visual 
features of dihadron and direct photon-hadron correlations. The dihadron 
correlations have two peaks at \dphi$\sim$~0 and $\pi$ corresponding to 
intrajet and back-to-back jet correlations, respectively. The isolated 
direct photon-hadron correlations show away-side yields that are 
consistently smaller than the corresponding \pion-hadron yields, which 
is to be expected because the \pion-hadron correlations probe larger hard 
scales due to the \pion being a fragment of the jet. Note that a 9\% 
charged hadron normalization uncertainty is not explicitly shown on any 
of figures displaying the per-trigger yields; this is denoted in each of 
the captions where the uncertainty applies.

\subsection{\pout distributions}

\begin{figure}[tbh]
	\includegraphics[width=1.0\linewidth]{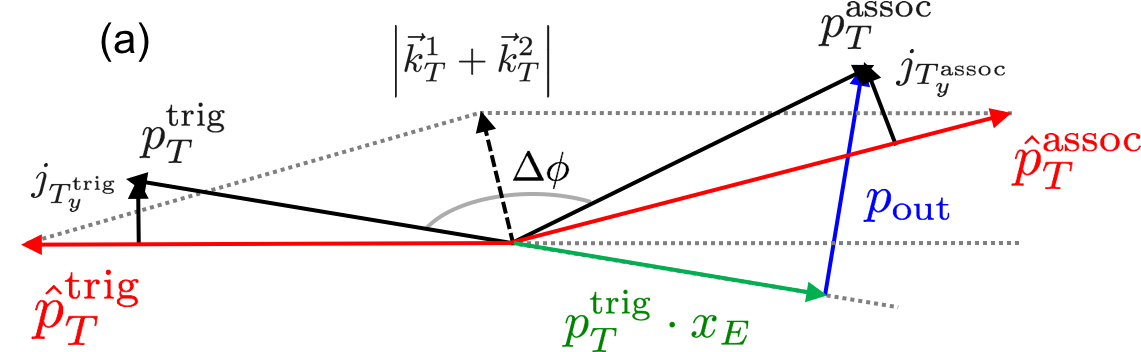}
    \includegraphics[width=1.0\linewidth]{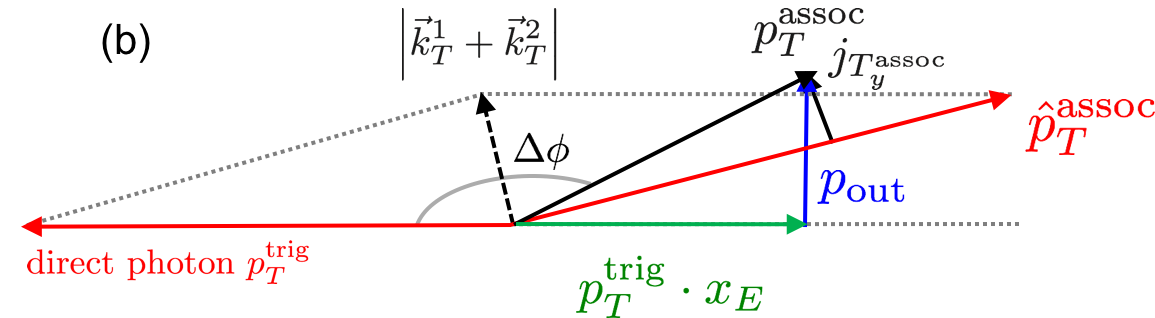}
    \caption{A diagram which shows the hard scattering kinematics of a 
nearly back-to-back correlation for (a) dihadron and (b) direct 
photon-hadron events, adapted from Ref.~\cite{Adare:2016bug}. Two 
hard-scattered partons, shown in red, are acoplanar due to the 
initial-state $\vec{k}_T^1$ and $\vec{k}_T^2$ of the colliding partons. 
The partons result in a trigger and associated jet fragment \pttrig and 
\ptassoc with a transverse momentum component perpendicular to the jet 
axis $j_{T^{\rm trig}_y}$ and $j_{T^{\rm assoc}_y}$ in the transverse 
plane, which are assumed to be Gaussian such that $\jt=\sqrt{2\langle 
j^{2}_{T^{\rm trig}_y}\rangle} =\sqrt{2\langle j^{2}_{T_y^{\rm 
assoc}}\rangle}$. For direct photons (b) \pttrig corresponds to the hard 
scattering vector because the direct photon is produced from the hard 
scattering. In each figure the quantity $x_E$ is labeled as the green 
vector and approximates the momentum fraction $z$ of the final-state 
away-side hadron.}
    \label{fig:ktkinematics}
\end{figure}

The hard scattering quantity $\pout=\ptassoc\sin\dphi$, the transverse 
momentum component of the associated hadron with respect to the trigger 
particle axis, is defined kinematically in Fig.~\ref{fig:ktkinematics}. 
Rather than constructing the per-trigger yields as a function of \dphi, 
they are instead constructed as a function of \pout in a similar way to 
the \dphi correlations. These distributions are the quantity of interest 
because the nonperturbative TMD structure can be observed from the 
correlation functions; additionally they have the advantage that the 
nonperturbative component of the away-side jet width can be separated 
from the perturbative component in momentum space with a transition 
region between the two components.

Reference~\cite{Adare:2016bug} found that the nonperturbative momentum 
widths of the \pout distributions binned in a fixed range of \ptassoc 
decreased with the hard scale \pttrig, contrary to what is qualitatively 
expected from CSS evolution. The \pout distributions were binned in a 
fixed \ptassoc range only, which means that the longitudinal momentum 
fraction $z$ of the away-side hadron with respect to the away-side jet 
was generally decreasing as the hard scale increased. Here we bin the 
\pout correlation distributions as a function of the quantity $x_E$, 
which is defined in Ref.~\cite{Adare:2010yw} as

\begin{equation}\label{eq:xe}
	x_E\equiv-\frac{\pttrig\cdot\ptassoc}{|\pttrig|^2} = -\frac{|\ptassoc|}{|\pttrig|}\cos\dphi
\end{equation}

\noindent and is geometrically shown in Fig.~\ref{fig:ktkinematics}. 
Because full jet reconstruction within PHENIX severely limits available 
statistics due to the limited acceptance, this quantity can be used as a 
proxy for $z$, the longitudinal momentum fraction of the associated 
away-side hadron with respect to the away-side jet. Although $x_E$ is an 
approximation for $z$, this alternative binning allows for a clearer 
comparison of the \pout distributions between hard scales as the 
associated hadrons are then binned in a similar kinematic way that is 
normalized by the near-side \pttrig.

Because \xe is only a proxy for $z$, there is an embedded dependence on 
the correlation between these two quantities for both the \pion-hadron 
and direct photon-hadron correlations. Previous correlations 
measurements from PHENIX have shown that the quantity 
$\hat{x}_h=\hat{p}_T^{\rm assoc}/\hat{p}_T^{\rm trig}$, where quantities 
with a hat indicate partonic quantities, is on average less than 
unity~\cite{Adare:2010yw}, and this has also been shown in direct 
photon-jet correlations~\cite{Sirunyan:2017qhf}. Thus for direct-photon 
hadron correlations, $z>\xe$. For \pion-hadron correlations, there is an 
additional dependence on $\langle z_T\rangle=\pttrig/\hat{p}_T^{\rm 
trig}$, which is roughly 0.6 at Relativistic-Heavy-Ion-Collider (RHIC) 
energies~\cite{Adler:2006sc}. Thus, for \pion-hadron correlations, on 
average $z<\xe$; therefore the dihadron and direct photon-hadron 
correlations are on average probing different values of the away-side 
hadron momentum fraction $z$.

\begin{figure}[tbh]
	\includegraphics[width=1.0\linewidth]{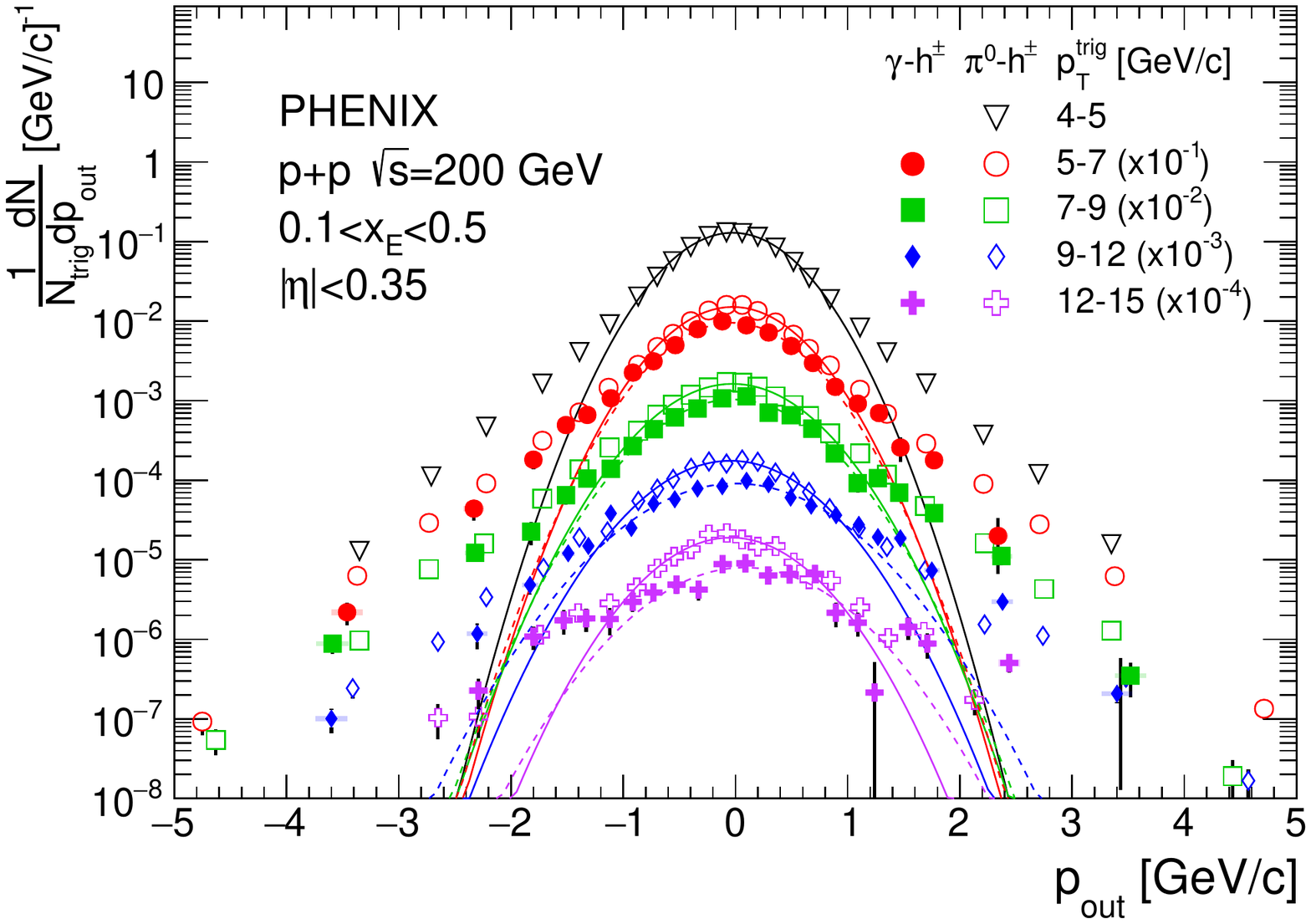}
    \caption{The \pout distributions are shown for dihadron and direct 
photon-hadron correlations, binned in $x_E$. The 9\% charged hadron 
normalization uncertainty is not explicitly shown on the figure.}
    \label{fig:pout_run15}
\end{figure}

The \pout distributions for \pp collisions at \sqs=~200 GeV binned in 
$0.1<x_E<0.5$ are shown in Fig.~\ref{fig:pout_run15}. The open points 
are the \pion-hadron correlations, while the filled points are the 
direct photon-hadron correlations. In constructing the correlations, the 
underlying event was statistically subtracted following a method similar 
to Ref.~\cite{Adare:2016bug}. The functions used to statistically 
subtract the underlying event are shown as fits to the away-side 
distributions in Fig.~\ref{fig:deltaphi_correlations}. Although the 
correlation functions are binned in $x_E$ instead of \ptassoc, there is 
still a clear transition from nonperturbative to perturbative 
sensitivity in the distributions. This is highlighted by the Gaussian 
fits to the small \pout region [-1.1,1.1] \gevc in the figure, where the 
fits clearly fail at describing the correlations at large \pout. We also 
note that Ref.~\cite{Adare:2016bug} found that the large \pout region 
was described reasonably well with a Kaplan fit; here the distributions 
are not described by a Kaplan fit due to the smaller center-of-mass 
energy. When \sqs is smaller, it is less likely that a high \pt gluon 
radiation will occur such that \pout is large. This causes the \pout 
distributions to fall more quickly towards zero at large \pout.

\begin{figure}[tbh]
	\includegraphics[width=1.0\linewidth]{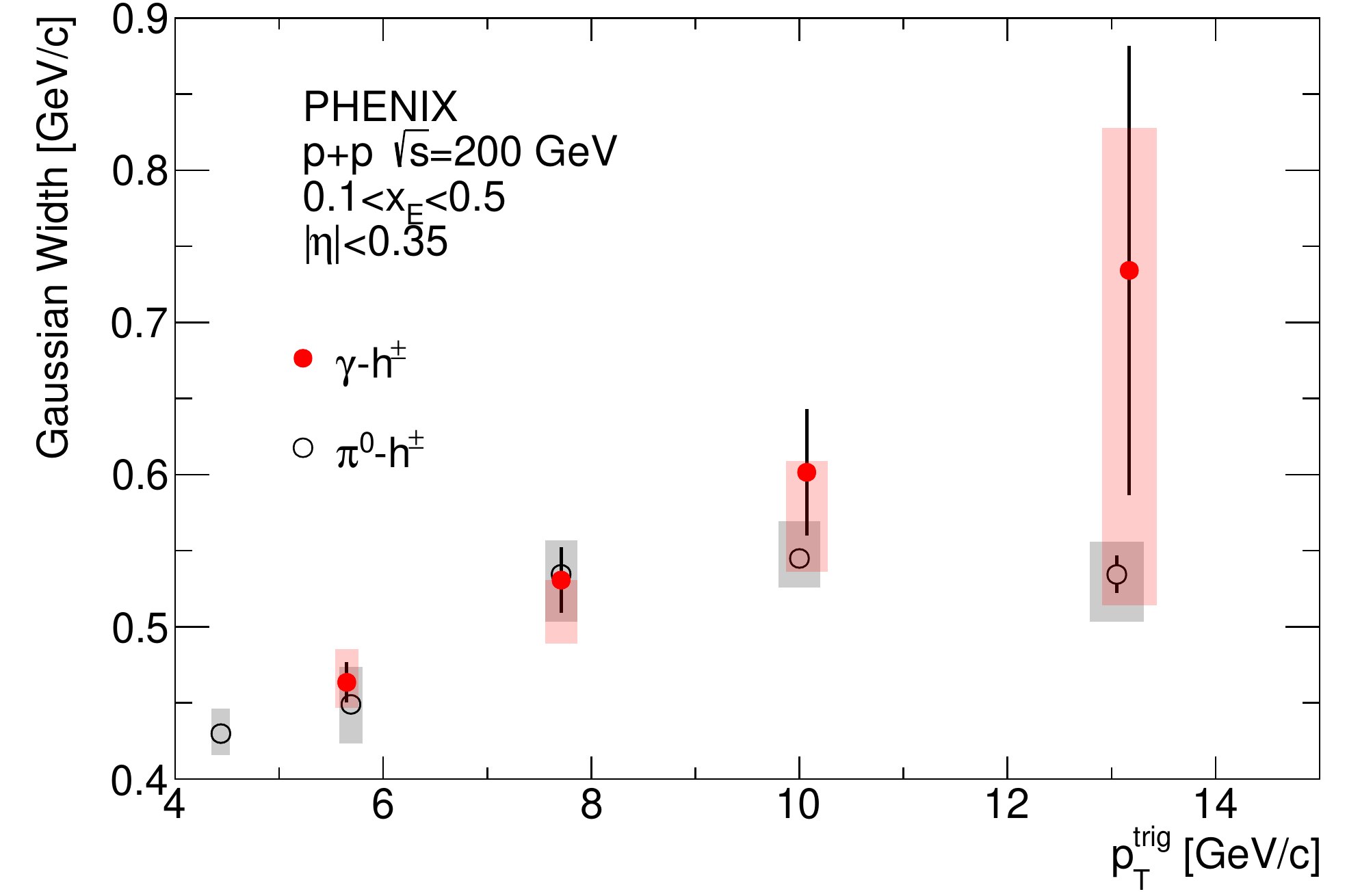}
    \caption{The nonperturbative away-side jet widths as a function of 
\pttrig at \sqs=~200 GeV for both direct photon-hadron and dihadron 
correlations.}
    \label{fig:run15_gausswidths}
\end{figure}

The widths of the Gaussian fits are extracted to quantify the evolution 
of the nonperturbative away-side jet widths as a function of \pttrig. 
Systematic uncertainties are evaluated by adjusting the fit region by 
$\pm$~0.2 \gevc and taking the absolute difference of the resulting 
Gaussian width. The values are shown in Fig.~\ref{fig:run15_gausswidths} 
and Table~\ref{tab:widths_200gev} and clearly demonstrate that the 
widths increase with \pttrig. This is in contrast to 
Ref.~\cite{Adare:2016bug}, where the widths decreased as a function of 
\pttrig when the \pout distributions were binned in a fixed \ptassoc 
range and thus not in a way to account for the longitudinal momentum 
fraction of the away-side hadron with respect to the near-side trigger 
particle.

 \begin{table}[tbh]
 \caption{\label{tab:widths_200gev}
    Gaussian widths of \pout for direct photon-hadron and dihadron 
correlations in \pp collisions at \sqs=~200 GeV. Units are [\gevc] for 
both $\langle\pttrig\rangle$ and the Gaussian widths.
}
 \begin{ruledtabular} \begin{tabular}{ccccc}
  Trigger Type & $\langle\pttrig\rangle$ & Gaussian Width &  Stat. & Sys. \\\hline
  $\pion$ & 4.44 & 0.429 & 0.001 & $^{+0.016}_{-0.014}$ \\
          & 5.69 & 0.449 & 0.001 & $^{+0.024}_{-0.026}$ \\
          & 7.71 & 0.534 & 0.002 & $^{+0.022}_{-0.031}$ \\
          & 10.1 & 0.545 & 0.004 & $^{+0.024}_{-0.018}$ \\ 
          & 13.1 & 0.534 & 0.012 & $^{+0.021}_{-0.031}$ \\
\\
 Direct photon & 5.65 & 0.46 & 0.01 & $^{+0.02}_{-0.02}$ \\
               & 7.71 & 0.53 & 0.02 & $^{+0.01}_{-0.04}$ \\
               & 10.1 & 0.60 & 0.04 & $^{+0.01}_{-0.06}$ \\
               & 13.2 & 0.73 & 0.15 & $^{+0.09}_{-0.22}$ \\
 \end{tabular} \end{ruledtabular}
 \end{table}

\begin{figure}[tbh]
    \includegraphics[width=1.0\linewidth]{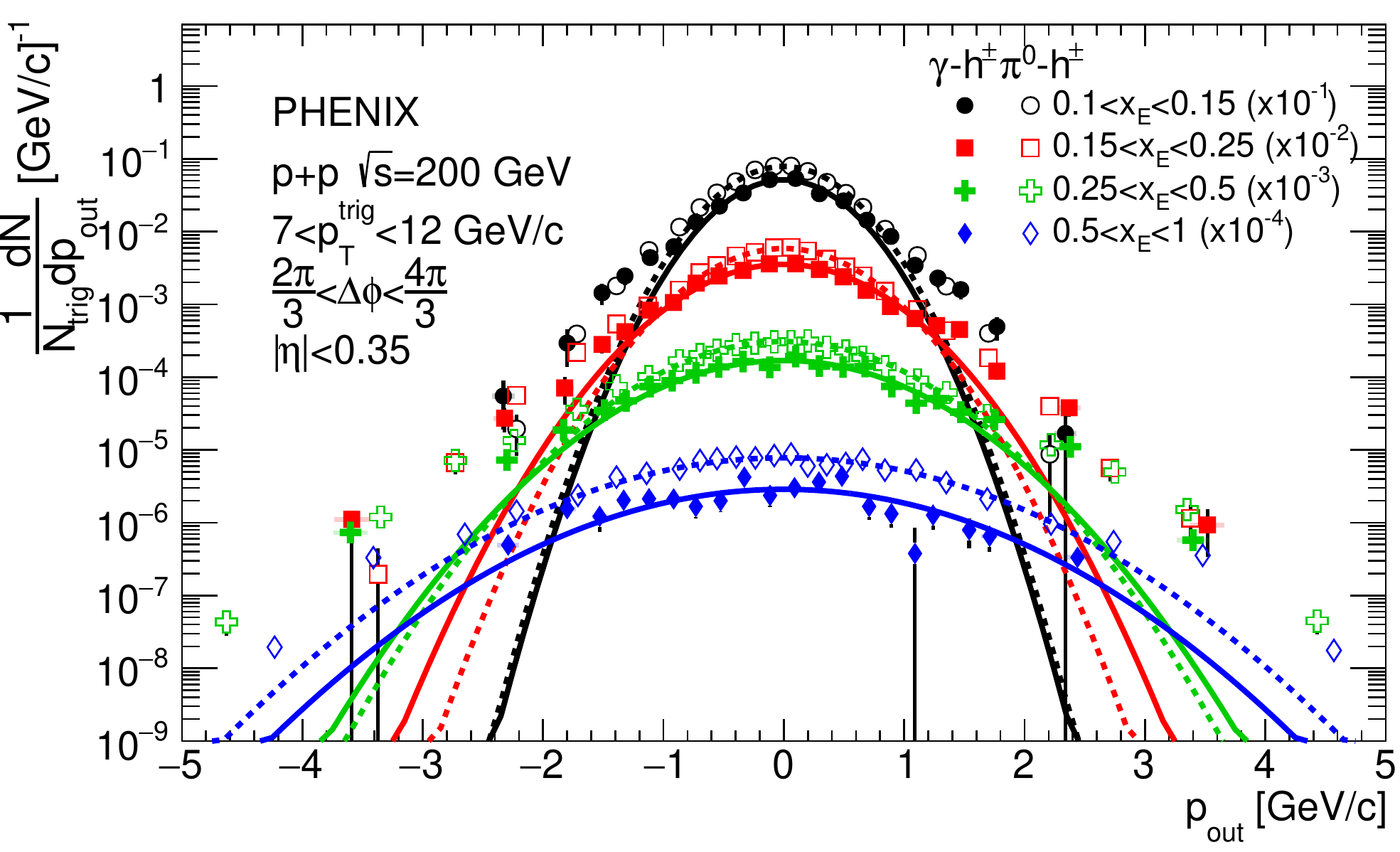}
    \caption{The \pout distributions are shown in several bins of \xe, 
integrated over a range of \pttrig. The 9\% charged hadron normalization 
uncertainty is not explicitly shown on the figure.}
    \label{fig:pout_dist_xe}
\end{figure}

\begin{figure}[tbh]
    \includegraphics[width=1.0\linewidth]{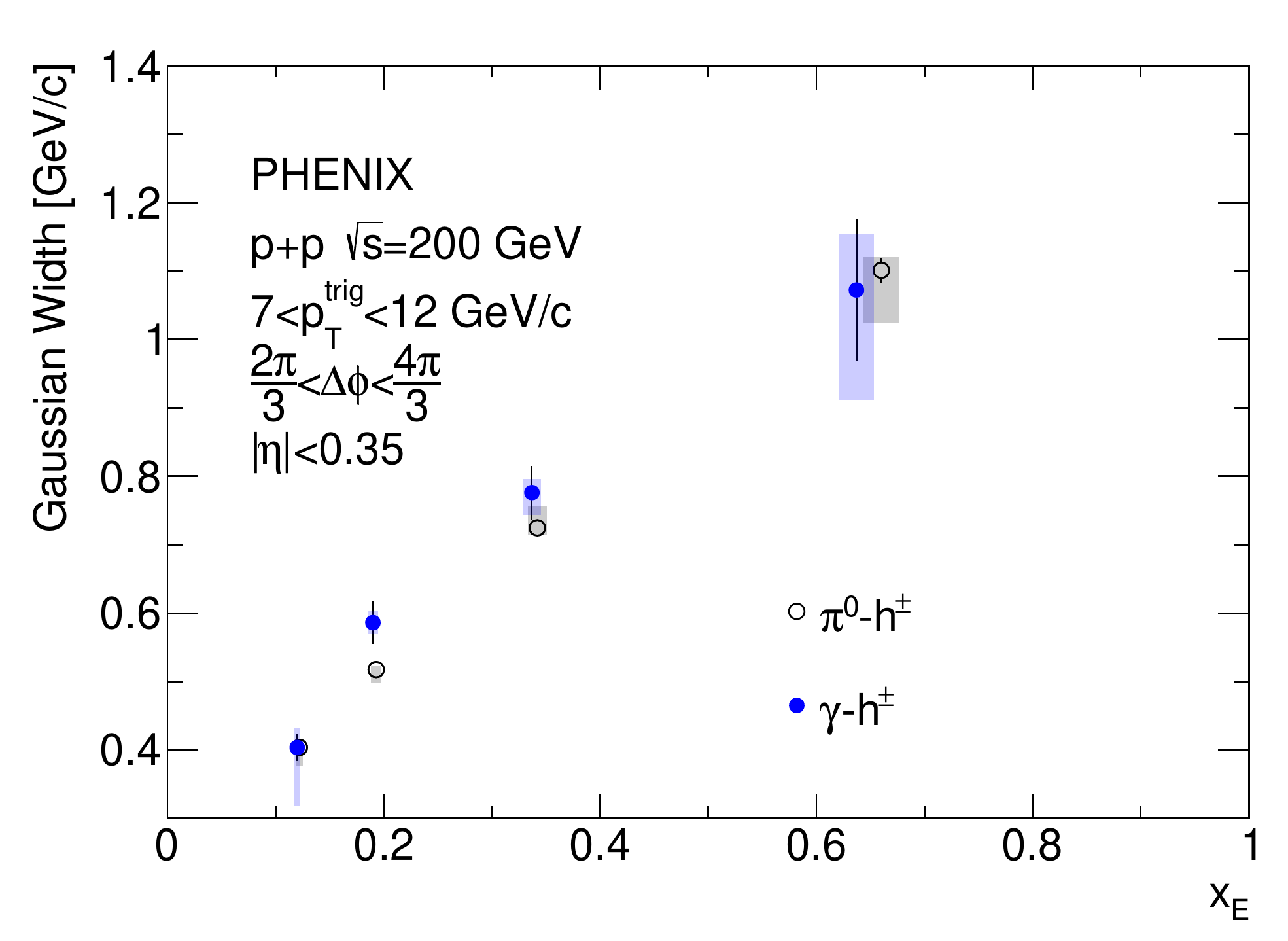}
    \caption{The Gaussian widths of \pout as a function of \xe are shown 
for both \pion and direct photon triggered correlations.}
    \label{fig:gausswidth_fxn_xe}
 \end{figure}

To study the dependence of the nonperturbative momentum widths on the 
fragmentation quantity \xe, the \pout distributions were constructed as 
a function of \xe. To have sufficient statistical precision, and to 
compare to the \sqs=~510 GeV data which is described later, the 
distributions were integrated over a larger range of $7<\pttrig<12$ 
\gevc. The \pout distributions are shown in Fig.~\ref{fig:pout_dist_xe} 
and the nonperturbative structure is fit with a Gaussian function, shown 
as dashed lines and solid lines for \pion and direct photon triggered 
correlations, respectively. The nonperturbative to perturbative 
transition is still visible; however, the region of both large \pout and 
\xe lacks statistical precision. The Gaussian fits are performed in 
varying regions of \pout, depending on the \xe bin because the 
nonperturbative structure strongly depends on the \xe bin probed. The 
Gaussian widths are extracted and shown as a function of \xe in 
Fig.~\ref{fig:gausswidth_fxn_xe}, where the systematic uncertainties on 
the widths are estimated in a similar way to the previous 
nonperturbative momentum widths.

\begin{figure}[tbh]
	\includegraphics[width=1.0\linewidth]{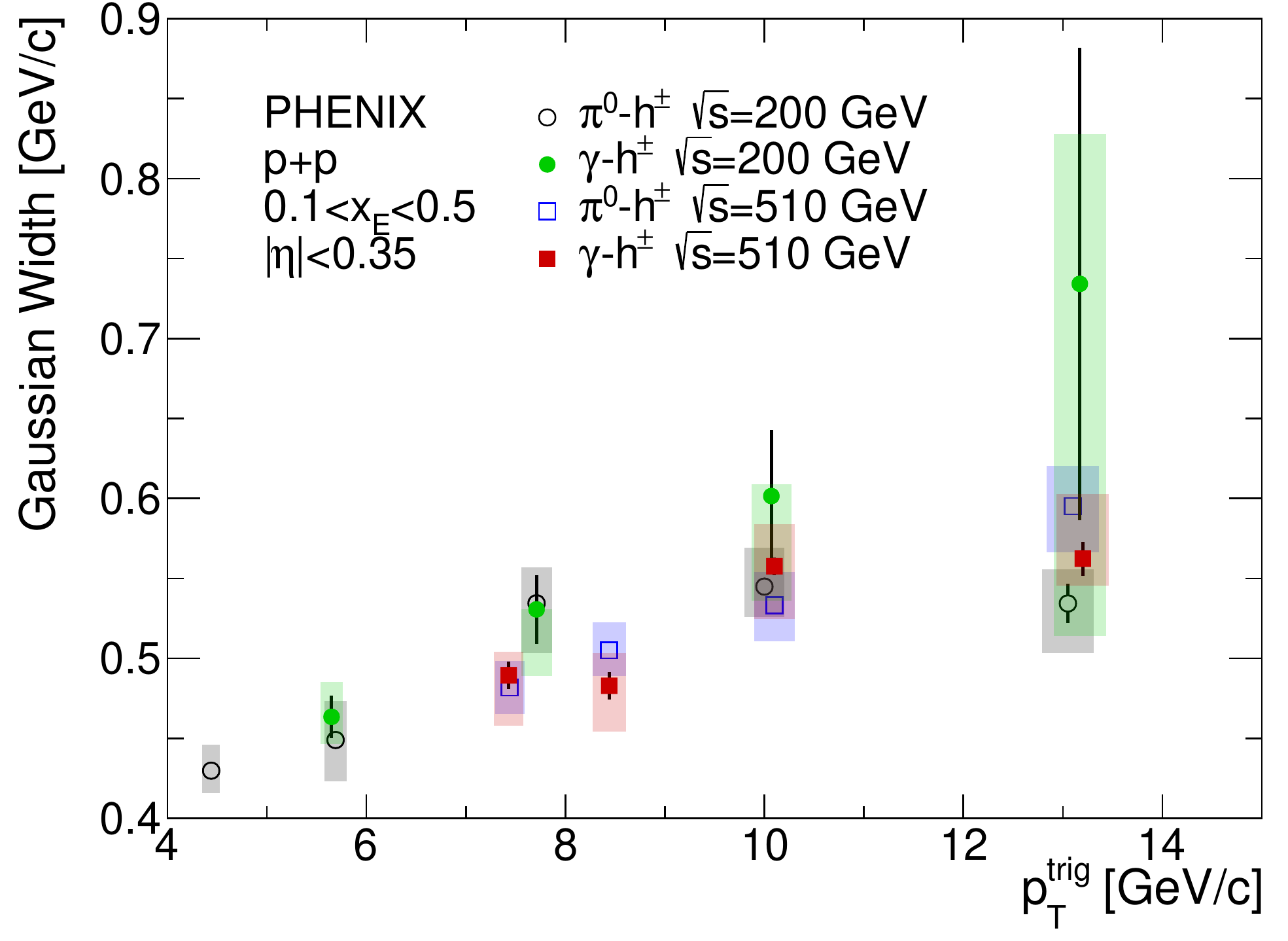}
    \caption{The Gaussian widths extracted from the \pout distributions 
in both \sqs=~200 GeV and \sqs=~510 GeV are shown as a function of 
\pttrig.}
    \label{fig:pt_gausswidths}
\end{figure}

\begin{figure}[tbh]
	\includegraphics[width=1.0\linewidth]{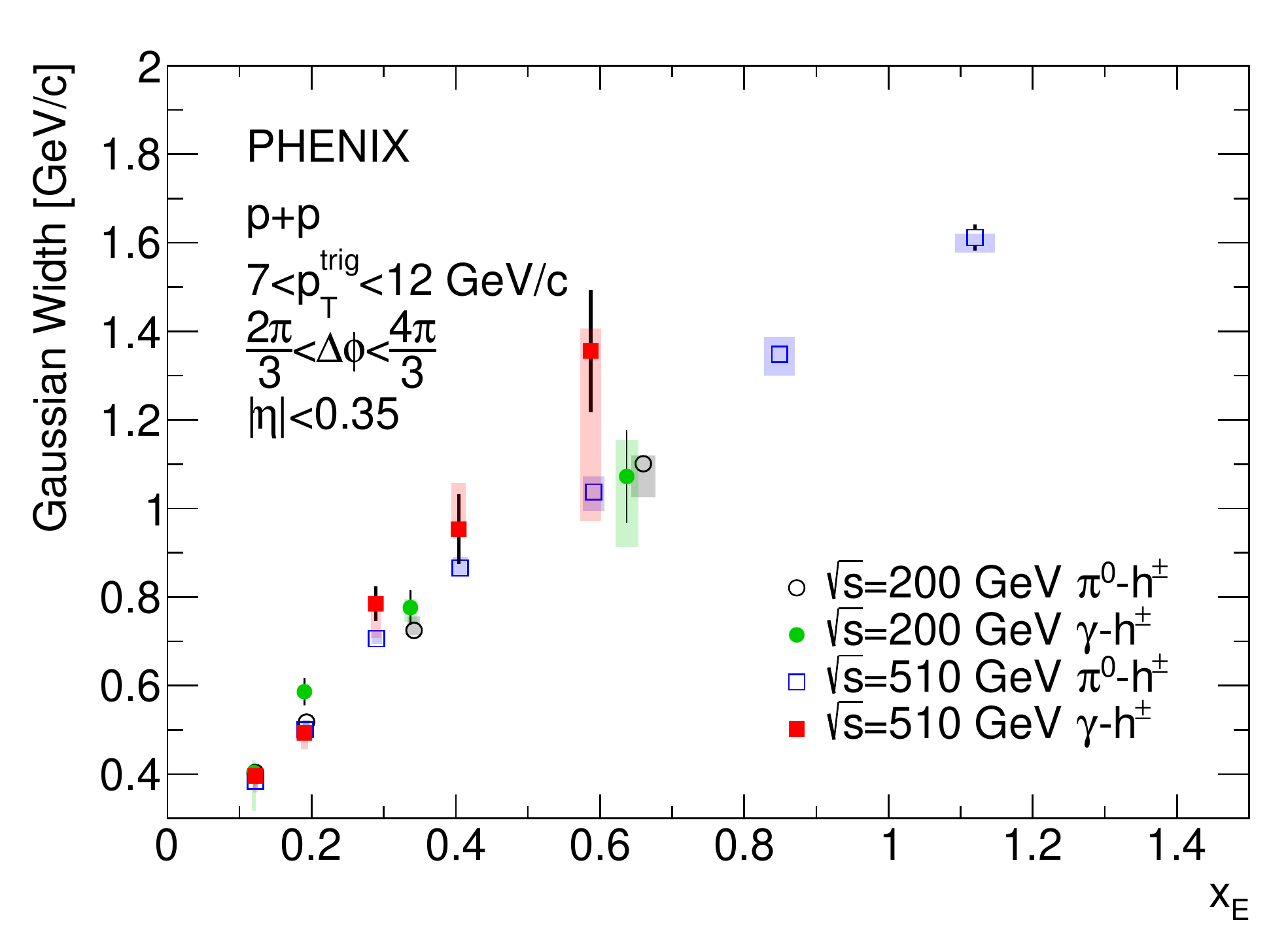}
    \caption{The nonperturbative Gaussian widths are shown as a function 
of \xe at two different center-of-mass energies.}
    \label{fig:gausswidth_fxn_xe_sqrts}
\end{figure}

To compare the results measured here to the previous PHENIX data at 
\sqs=~510 GeV, the data from Ref.~\cite{Adare:2016bug} were rebinned in 
$x_E$ similarly to the results shown here. The nonperturbative momentum 
widths are extracted from Gaussian fits to the small \pout region and 
the widths from both center-of-mass energies are shown as a function of 
\pttrig in Fig.~\ref{fig:pt_gausswidths}. Note that only the 
\pion-hadron correlations with \pttrig$>$7 \gevc were reanalyzed; this 
is because, due to detector capabilities, there was a minimum \ptassoc 
limit of 0.7 \gevc in the \sqs=~510 GeV analysis~\cite{Adare:2016bug}. 
For the minimum $x_E$ cut of 0.1 to be unbiased, we require that the 
\pttrig$>$7 \gevc in the \sqs=~510 GeV data. The values at the two 
different center-of-mass energies are consistent with one another at a 
similar \pttrig within uncertainties. 
Figure~\ref{fig:gausswidth_fxn_xe_sqrts} shows the nonperturbative 
Gaussian widths as a function of \xe at the two different center-of-mass 
energies. The nonperturbative Gaussian widths also show little 
dependence as a function of \xe on the center-of-mass energy. The data 
shown here will provide additional constraints on processes predicted to 
break TMD factorization and appear to show that, at the hard scales and 
energies probed by RHIC, the nonperturbative away-side widths are 
consistent within uncertainties at two different center-of-mass 
energies. Similar conclusions have been drawn by the STAR collaboration, 
where polarized TMD observables are consistent within uncertainties 
between \sqs=~200 and 500 GeV~\cite{Adamczyk:2017wld,Adamczyk:2017ynk}.

\begin{figure}[tbh]
	\includegraphics[width=1.0\linewidth]{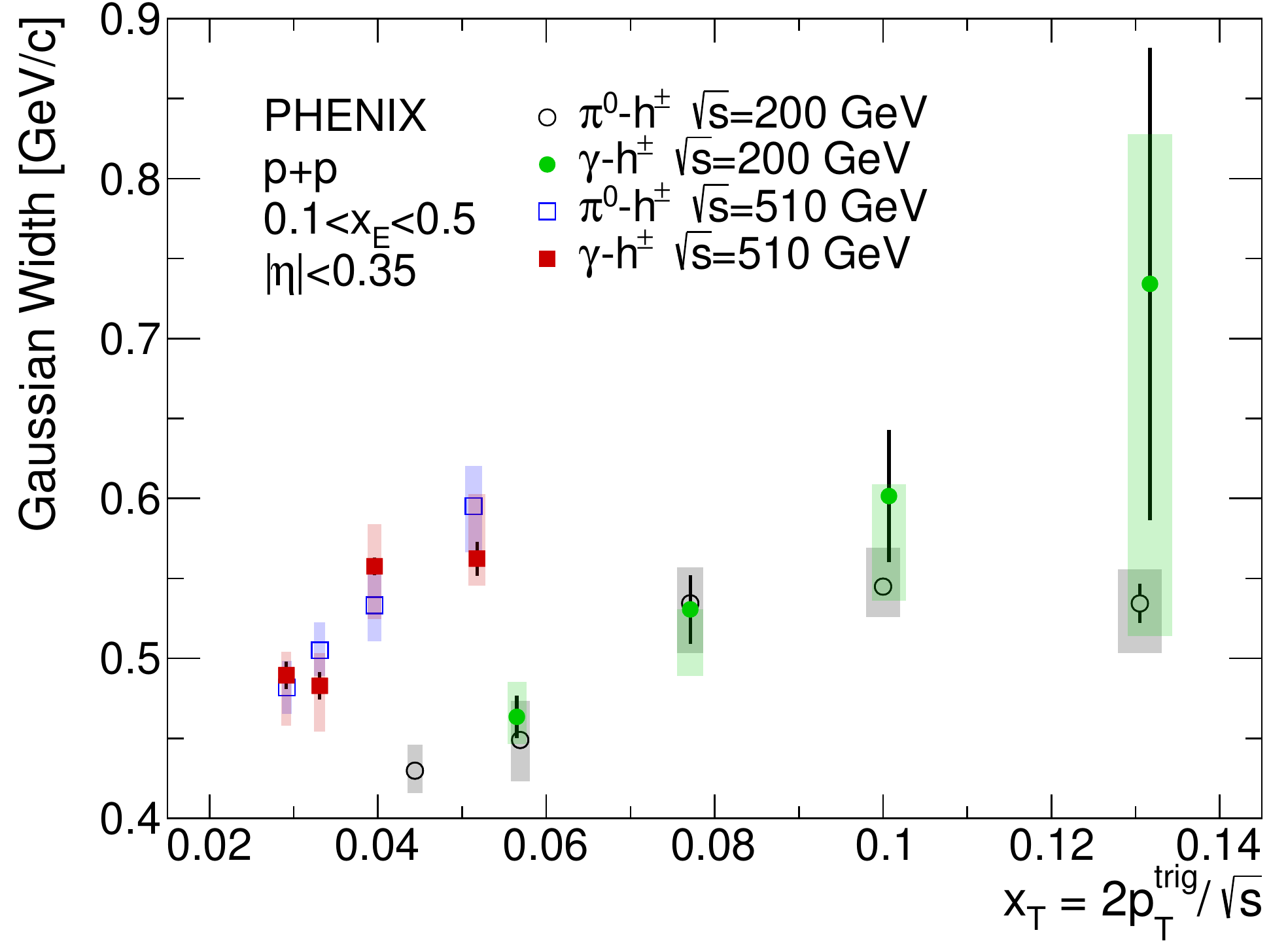}
    \caption{The Gaussian widths extracted from the \pout distributions 
in both \sqs=~200 GeV and \sqs=~510 GeV are shown as a function of 
$x_T=2\pttrig/\sqs$.}
    \label{fig:xt_gausswidths}
\end{figure}

The nonperturbative away-side jet widths are also shown as a function of 
$x_T=2\pttrig/\sqs$ in Fig.~\ref{fig:xt_gausswidths}.  The Gaussian 
widths do not appear to scale with $x_T$; however, they appear to show 
qualitatively similar behavior to DY interactions. The nonperturbative 
momentum widths sensitive to a small transverse momentum scale increase 
with \sqs at a similar $x_T$. This behavior as a function of 
$\sqrt{\tau}=x_1x_2$, where $x_1$ and $x_2$ are the partonic momentum 
fractions of the quark antiquark pair, and \sqs can be observed in TMD 
momentum widths measured from DY data (see e.g.~\cite{Ito:1980ev}). 
However, it is interesting to note that in DY at similar $M_{\mu\mu}$ 
nonperturbative momentum widths clearly rise with \sqs, while in the 
measurements presented here, as well as other polarized TMD observables 
from RHIC~\cite{Adamczyk:2017ynk,Adamczyk:2017wld}, nonperturbative 
momentum widths are consistent with each other as a function of \pttrig 
and \sqs.

 \begin{table}[tbh]
 \caption{\label{tab:widths_510gev}
    Gaussian widths of \pout for direct photon-hadron and dihadron 
correlations in \pp collisions at \sqs=~510 GeV in a fixed $x_E$ bin, as 
seen in Fig.~\ref{fig:xt_gausswidths} and rebinned from 
Ref.~\cite{Adare:2016bug}. Units are [\gevc] for both 
$\langle\pttrig\rangle$ and the Gaussian widths.
 }
 \begin{ruledtabular}
 \begin{tabular}{ccccc}
  Trigger Type & $\langle\pttrig\rangle$ & Gaussian Width &  Stat. & Sys. \\\hline
  $\pion$ & 7.44 & 0.482 & 0.001 & $^{+0.016}_{-0.017}$ \\
          & 8.44 & 0.505 & 0.001 & $^{+0.018}_{-0.016}$ \\
          & 10.1 & 0.533 & 0.001 & $^{+0.021}_{-0.022}$ \\
          & 13.1 & 0.595 & 0.001 & $^{+0.025}_{-0.029}$ \\
\\
 Direct photon & 7.43 & 0.489 & 0.009 & $^{+0.015}_{-0.032}$ \\
               & 8.44 & 0.483 & 0.009 & $^{+0.021}_{-0.029}$ \\
               & 10.1 & 0.558 & 0.007 & $^{+0.026}_{-0.033}$ \\
               & 13.2 & 0.562 & 0.011 & $^{+0.040}_{-0.017}$ \\
 \end{tabular}
 \end{ruledtabular}
 \end{table}

\section{Conclusions}

Dihadron and direct photon-hadron correlations were measured at the 
PHENIX experiment in \pp collisions at \sqs=~200 GeV. These processes 
have been predicted to violate factorization in a TMD framework due to 
soft gluon exchanges with remnants that are possible in both the initial 
and final states in hadronic 
collisions~\cite{Collins:2007nk,Rogers:2010dm}. To better interpret the 
possible factorization breaking effects that were studied in 
Ref.~\cite{Adare:2016bug}, the \pout distributions were binned in the 
hard scattering variable $x_E$ to provide more control over the 
fragmentation dependence as a function of the hard scale of the 
interaction. When accounting for the away-side hadron \ptassoc with 
respect to the near-side \pttrig, the nonperturbative momentum widths 
are found to increase with the hard scale of the interaction \pttrig. 
This is qualitatively similar to DY interactions in that nonperturbative 
momentum widths sensitive to a small transverse momentum scale increase 
with the hard scale of the interaction. This behavior has been verified 
phenomenologically in both DY and SIDIS 
processes~\cite{Landry:2002ix,Konychev:2005iy,Schweitzer:2010tt,Aidala:2014hva,Nadolsky:1999kb}, 
and this measurement shows that this behavior is similar in processes 
which are predicted to break factorization.

For the time being, the only obvious way to quantify factorization 
breaking effects is to compare data in processes that are predicted to 
violate factorization and calculations assuming factorization holds. 
Such calculations are not available at this time, largely due to the 
fact that even for unpolarized observables TMD PDFs and TMD FFs are not 
yet well constrained due to a lack of data. The first global fit of TMD 
data was very recently reported~\cite{Bacchetta:2017gcc}, and in 
particular the study calls for more data covering a broader range of 
kinematic variables to constrain future global fits. Future comparisons 
of magnitudes, shapes, and evolution of these observables to 
calculations should be made to quantify the magnitude of factorization 
breaking effects; for example, the rate at which TMD observables evolve 
in processes predicted and not predicted to break factorization may be 
different.

Nonetheless, additional observables can be identified to provide more 
data to quantify effects from factorization breaking. For example, spin 
asymmetries in the direct photon-jet channel have been predicted to 
arise due to factorization breaking effects~\cite{Rogers:2013zha}; 
polarized $p$$+$$A$ collisions have also been proposed as an avenue to 
quantify factorization breaking effects in the same hard scattering 
channel~\cite{Schafer:2014xpa}. Unpolarized processes still have the 
potential to quantify factorization breaking effects; for example, 
nearly back-to-back direct photon-quarkonium production could be used to 
quantify factorization breaking effects as well, depending on whether or 
not the quarkonium state is produced in a color singlet or color octet 
state. Measuring processes that may be sensitive to factorization 
breaking effects, including those presented here, will be important for 
better understanding the characteristics of QCD as a non-Abelian 
gauge-invariant field theory. The exploration of the role of color in 
hadronic interactions is deeply connected to the unique properties of 
QCD, and future measurements will continue to constrain the magnitudes 
of these effects.

\section*{ACKNOWLEDGMENTS} 

We thank the staff of the Collider-Accelerator and Physics Departments 
at Brookhaven National Laboratory and the staff of the other PHENIX 
participating institutions for their vital contributions.  We also thank 
J. C. Collins for valuable discussions regarding the interpretation of 
these results.  
We acknowledge support from the Office of Nuclear Physics in the
Office of Science of the Department of Energy,
the National Science Foundation, 
Abilene Christian University Research Council, 
Research Foundation of SUNY, and
Dean of the College of Arts and Sciences, Vanderbilt University 
(U.S.A),
Ministry of Education, Culture, Sports, Science, and Technology
and the Japan Society for the Promotion of Science (Japan),
Conselho Nacional de Desenvolvimento Cient\'{\i}fico e
Tecnol{\'o}gico and Funda\c c{\~a}o de Amparo {\`a} Pesquisa do
Estado de S{\~a}o Paulo (Brazil),
Natural Science Foundation of China (People's Republic of China),
Croatian Science Foundation and
Ministry of Science and Education (Croatia),
Ministry of Education, Youth and Sports (Czech Republic),
Centre National de la Recherche Scientifique, Commissariat
{\`a} l'{\'E}nergie Atomique, and Institut National de Physique
Nucl{\'e}aire et de Physique des Particules (France),
Bundesministerium f\"ur Bildung und Forschung, Deutscher
Akademischer Austausch Dienst, and Alexander von Humboldt Stiftung (Germany),
J. Bolyai Research Scholarship, EFOP, the New National Excellence
Program ({\'U}NKP), NKFIH, and OTKA (Hungary),
Department of Atomic Energy and Department of Science and Technology (India), 
Israel Science Foundation (Israel), 
Basic Science Research Program through NRF of the Ministry of Education (Korea),
Physics Department, Lahore University of Management Sciences (Pakistan),
Ministry of Education and Science, Russian Academy of Sciences,
Federal Agency of Atomic Energy (Russia),
VR and Wallenberg Foundation (Sweden), 
the U.S. Civilian Research and Development Foundation for the
Independent States of the Former Soviet Union, 
the Hungarian American Enterprise Scholarship Fund,
the US-Hungarian Fulbright Foundation,
and the US-Israel Binational Science Foundation.


%
 
\end{document}